\documentclass[conference,a4paper]{IEEEtran}

\IEEEoverridecommandlockouts

\usepackage{cite}

\usepackage{graphicx}
\usepackage{subfigure}
\usepackage{boxedminipage}
\usepackage{comment}

\usepackage[cmex10]{amsmath}
\usepackage{amsmath}

\usepackage{algorithmic}

\usepackage{url}

\usepackage{balance}

\usepackage{flushend}
\flushend

\renewcommand{\P}{\mathrm{P}}
\newcommand{\E}{\mathrm{E}}

\newtheorem{theorem}{Theorem}

\usepackage{times,color}

\begin{document}
%
\title{Reliability-based Error Detection for Feedback Communication with Low Latency}

\author{\IEEEauthorblockN{Adam R. Williamson, Tsung-Yi Chen and Richard D. Wesel}
\IEEEauthorblockA{Department of Electrical Engineering,
University of California, Los Angeles, Los Angeles, California 90095\\
Email: adamroyce@ucla.edu; tychen@ee.ucla.edu; wesel@ee.ucla.edu}
\thanks{This research was supported by National Science Foundation Grant CIF CCF 1162501.}
}

\maketitle


%
%
\begin{abstract}
This paper presents a reliability-based decoding scheme for variable-length coding with feedback and demonstrates via simulation that it can achieve higher rates than Polyanskiy et al.'s  random coding lower bound for variable-length feedback (VLF) coding on both the BSC and AWGN channel.  The proposed scheme uses the reliability output Viterbi algorithm (ROVA) to compute the word error probability after each decoding attempt, which is compared against a target error threshold and used as a stopping criterion to terminate transmission. The only feedback required is a single bit for each decoding attempt, informing the transmitter whether the ROVA-computed word-error probability is sufficiently low. Furthermore, the ROVA determines whether transmission/decoding may be terminated without the need for a rate-reducing CRC.
\end{abstract}


\IEEEpeerreviewmaketitle

%
%
\section{Introduction}
\label{sec:intro}
Polyanskiy et al. \cite{Polyanskiy_CCR_2010} tightly characterized the back-off from capacity at short blocklengths without feedback by providing achievability and converse bounds on the maximum rate, along with a normal approximation using the channel dispersion (i.e., the stochastic variation of the channel) that closely approximates both bounds.
The results of  \cite{Polyanskiy_CCR_2010} show that there is a severe penalty in the maximum achievable rate for small blocklengths (on the order of several hundred bits or less).

In \cite{Polyanskiy_IT_2011_NonAsym}, Polyanskiy et al. investigated how including feedback improves the maximum achievable rate at short blocklengths.  Noiseless feedback does not increase the asymptotic (Shannon) capacity of memoryless channels \cite{Shannon_IT_1956}, but it {\em can} significantly improve the achievable rate at short blocklengths.  For (even the best) fixed-length block codes paired with an ARQ strategy, the maximum rate is slow to converge to the asymptotic capacity.  However, when {\em variable}-length coding is used on channels with noiseless feedback, the maximum rate can improve dramatically \cite{Polyanskiy_IT_2011_NonAsym} as compared to the no-feedback case for short average blocklength. 

Polyanskiy et al. considered two main categories of coding with feedback in  \cite{Polyanskiy_IT_2011_NonAsym}: variable-length feedback (VLF) coding, in which the receiver decides when to terminate transmission based on a desired probability of error $\epsilon$, and variable-length feedback coding with termination (VLFT), in which the receiver provides full, noiseless feedback to the transmitter, which uses an infinitely reliable feedforward channel to terminate the transmission when the receiver has decoded to the correct codeword.

In \cite{Williamson_ISIT_2012}, we presented a VLFT scheme based on a rate-compatible sphere-packing (RCSP) analysis and  tail-biting convolutional codes. We showed that when constrained to limited decoding attempts, VLFT implemented with rate-compatible, punctured, tail-biting convolutional codes can provide higher throughput than the random coding lower bound of \cite{Polyanskiy_IT_2011_NonAsym}.   

%
In this paper, we focus on VLF codes and demonstrate a scheme that achieves a higher rate than the random coding achievability bound of \cite{Polyanskiy_IT_2011_NonAsym}.  In particular, we show that trellis-based codes (e.g., convolutional codes) in which decoding is attempted after each received symbol using the reliability output Viterbi algorithm (ROVA) \cite{Raghavan_ROVA_TransIT_1998} can deliver rates surpassing the VLF random coding bound. 

The remainder of the paper proceeds as follows: Sec.~\ref{sec:VLF} reviews the VLF results of \cite{Polyanskiy_IT_2011_NonAsym}.  Sec.~\ref{sec:ROVA} presents the reliability output Viterbi algorithm (ROVA) and its use in the VLF setting.  Sec. \ref{sec:Simulation_Results} shows simulation results for VLF using convolutional codes and the ROVA, comparing these results with finite-blocklength bounds on the maximum rate for both the binary symmetric channel (BSC) and additive white Gaussian noise (AWGN) channel. Sec. \ref{sec:conc} concludes the paper.

%
%
\section{Variable-length Feedback (VLF) Codes}
\label{sec:VLF}

In the variable-length feedback (VLF) coding setting, the receiver attempts to decode the true message $W^*$ based on the sequence of $n$ symbols that have been received so far, $Y^n$. The receiver terminates the transmission when its estimate of the message $\hat{W}$ is sufficiently reliable to satisfy the $\epsilon$ error requirement, i.e., $\mathrm{P}(\hat{W} | Y^n) = \mathrm{P}(\hat X^n | Y^n) \geq {1 - \epsilon}$, where $\hat X^n$ is the transmitted sequence corresponding to message $\hat W$.

The transmitter is informed of the receiver's requests for additional coded symbols via noiseless ACK/NACKs, requiring only one bit of feedback per forward channel use (i.e., we are considering {\it stop-feedback codes} \cite{Polyanskiy_IT_2011_NonAsym}, or {\it decision feedback} \cite{Forney_Erasure_TransIT_1968}).
We denote average blocklength by $\ell$, the cardinality of the message alphabet $\mathcal{W}$ by $M$, and the required average probability of message error by $\epsilon$.

Polyanskiy et al.'s achievability result for VLF follows:
	\begin{theorem}[{\cite[Theorem 3]{Polyanskiy_IT_2011_NonAsym}}]
	For a scalar $\gamma > 0$, there exists an $(\ell, M, \epsilon)$ VLF code satisfying
	\begin{IEEEeqnarray}{rCl}
	\ell &\leq& \mathrm{E}[\tau], \\
	\epsilon &\leq& (M-1) \mathrm{P}[ \bar\tau \leq \tau],
	\end{IEEEeqnarray}
	where $\gamma$ is used as a threshold for determining the hitting times $\tau$ and $\bar\tau$:
	\begin{IEEEeqnarray}{rCl}
	\tau &=& \inf \{n \geq 0 : i(X^n;Y^n) \geq \gamma \} 	\label{eqn:vlf_tau} \\
	\bar\tau &=& \inf \{n \geq 0 : i(\bar X^n;Y^n) \geq \gamma \}.
	\end{IEEEeqnarray}
	\label{thm:achievability}
	\end{theorem}
\vspace{-14pt}
This lower bound uses random coding to prove that some code exists with at least the given cardinality $M$, though it doesn't specify how to efficiently compute the information densities $i(X^n;Y^n)$ for each of the $M$ codewords, for $n=1, 2, \dots$.

Polyanskiy et al.'s converse result for  VLF($\epsilon$) curves comes from \cite[Lemmas 1 and 2]{Burnashev_1976}:
	\begin{theorem}[{\cite[Theorem 4]{Polyanskiy_IT_2011_NonAsym}}]
	For an arbitrary DMC with capacity $C$ and $0 \leq \epsilon \leq 1 - \frac{1}{M}$, any $(\ell, M, \epsilon)$ VLF code satisfies:
	\begin{IEEEeqnarray}{rCl}
	\log M &\leq \frac{\ell C + h_\text{b}(\epsilon)}{1 - \epsilon}, 
	\end{IEEEeqnarray}
	where $h_\text{b}(\epsilon)$ is the binary entropy function.
	\label{thm:converse_awgn}
	\end{theorem}

When the maximal relative entropy $C_1$ is finite, we have the following converse result, which is tighter than Thm.~\ref{thm:converse_awgn}:
	\begin{theorem}[{\cite[Theorem 6]{Polyanskiy_IT_2011_NonAsym}}]
	For a DMC with $0$$<$$C$$\leq$$C_1$$<$$\infty$ and $0 < \epsilon \leq 1 - \frac{1}{M}$, any $(\ell, M, \epsilon)$ VLF code satisfies:
	\begin{IEEEeqnarray}{rCl}
	\ell &\geq \sup \limits_{0<\xi \leq 1 - \frac{1}{M}}  \left [ \frac{1}{C} A + B \right ],
	\end{IEEEeqnarray}
where
	\begin{IEEEeqnarray}{rCl}
	A &=& \log M - F_M(\xi) - \min \left (F_M(\epsilon), \frac{\epsilon}{\xi} \log M \right ), \\
	B &=& 	\left| \frac{1-\epsilon}{C_1} \log \frac{\lambda_1 \xi}{\epsilon (1-\xi)} - \frac{h_b(\epsilon)}{C_1}  \right |^{+},  \\
	C_1 &=& \max \limits_{x_1,x_2} D \left(\mathrm{P}_{Y|X=x_1}|\mathrm{P}_{Y|X=x_2} \right), \\
	F_M(x) &=& x \log (M-1) + h_b(x), 0 \leq x \leq 1, \\
	\lambda_1 &=& \min \limits_{y,x_1,x_2} \frac {\mathrm{P}_{Y|X}(y|x_1)} {\mathrm{P}_{Y|X}(y|x_2)} \in (0,1).
	\end{IEEEeqnarray}
	\label{thm:converse_bsc}
	\end{theorem}

As shown in Fig.~\ref{fig:ROVA_bsc_sims} for the binary symmetric channel with crossover probability $p$=$0.05$, there is a considerable gap between the lower (achievability) and upper (converse) bounds on the maximum rate at short blocklengths. In this paper we demonstrate that convolutional codes can deliver performance surpassing the lower bound, suggesting that other bounding techniques besides those based on random coding may be appropriate for especially short blocklengths.

%
%
\section{The Reliability Output Viterbi Algorithm}
\label{sec:ROVA}

In order to use VLF codes in practice, the receiver must compute the probability that the codeword $\hat X^n$ of length $n$ it has decoded is correct, $\mathrm{P}(\hat X^n|Y^n)$. This can be calculated as
\begin{IEEEeqnarray}{rCl}
	\mathrm{P}(\hat X^n|Y^n) &=& \frac{\mathrm{P}(Y^n|\hat X^n) \mathrm{P}(\hat X^n)}{\mathrm{P}(Y^n)} \\
	&=& \frac{\mathrm{P}(Y^n|\hat X^n) \mathrm{P}(\hat X^n)}{\sum \limits_{m'} \mathrm{P}(Y^n|X_{m'}^n) \mathrm{P}(X_{m'}^n)},
\end{IEEEeqnarray}
which can be further simplified if each of the codewords $X_{m'}^n$ is {\it a priori} equally likely, i.e., $\mathrm{P}(X_{m}^n) = \mathrm{P}(X_{m'}^n) ~\forall m \neq m'$. This yields
\begin{IEEEeqnarray}{rCl}
	\mathrm{P}(\hat X^n|Y^n) &= \frac{\mathrm{P}(Y^n|\hat X^n) }{\sum \limits_{m'} \mathrm{P}(Y^n|X_{m'}^n) }.
	\label{eqn:P_x_y}
\end{IEEEeqnarray}
In general, the denominator in \eqref{eqn:P_x_y} may be computationally intractable when the message set cardinality $M$ is large. However, in much the same way that the Viterbi algorithm takes advantage of the trellis structure of convolutional codes to compute $\mathrm{P}(Y^n | \hat X^n)$ for maximum-likelihood decoding, an augmented Viterbi decoder referred to as the reliability output Viterbi algorithm (ROVA) \cite{Raghavan_ROVA_TransIT_1998} can be used to compute $\mathrm{P}(\hat X^n|Y^n)$ exactly with relative efficiency. The complexity of the ROVA is linear in the blocklength and exponential in the constraint length of the code (i.e., it has complexity on the same order as that of the original Viterbi algorithm). This probability can also be computed approximately by the simplified (approximate) ROVA \cite{Fricke_Approx_ROVA_VTC_2007}.

The ROVA can be used to compute the probability of word error for any finite-state Markov process observed via memoryless channels (e.g., in maximum-likelihood sequence estimation for signal processing applications).


%
%
\subsection{Reliability-based Retransmissions}

In \cite{Fricke_Reliability_HARQ_TCOM_2009}, Fricke et al. proposed a reliability-based retransmission criteria for hybrid ARQ, using the word error probability calculated by the decoder. Similarly, in order to investigate the maximum rate at short blocklengths, we use the ROVA to compute $\mathrm{P}(\hat X^n|Y^n)$. If the computed word error probability is larger than the target error probability $\epsilon$, our decoder signals that additional codeword symbols are required (e.g., sends a NACK), and the transmitter sends another coded symbol. When $\mathrm{P}(\hat X^n|Y^n)$ is computed exactly, this scheme guarantees that the overall undetected error rate will converge to a value less than the target $\epsilon$.

 As compared to setting aside some parity bits to be used only for error detection (e.g., in a CRC), a reliability-based feedback approach allows all coded symbols to be used for both error correction and detection, improving throughput.

\subsection{Relationship to Generalized Decoding}

Hof et al. \cite{Hof_Conv_Bounds_ISCTA_2009} provides a modification to the Viterbi algorithm which permits erasure decoding according to Forney's generalized decoding rule \cite{Forney_Erasure_TransIT_1968}, based upon a predetermined erasure threshold $T$. Using Forney's rule, a decoder picks codeword $X_m^n$ if the following is satisfied:
\begin{IEEEeqnarray}{rCl}
	\frac{\mathrm{P}(Y^n,X_m^n)}{\sum_{m' \neq m} \mathrm{P}(Y^n,X_{m'}^n)} \geq e^{nT},
\end{IEEEeqnarray}
which is equivalent to the following condition:
\begin{IEEEeqnarray}{rCl}
	\mathrm{P}(X_m^n|Y^n) \geq \frac{e^{nT}}{1+e^{nT}}.
\end{IEEEeqnarray}

We note that setting the threshold $T$ according to $nT~=~\log~\frac{1-\epsilon}{\epsilon}$ coincides with the ROVA, in which a target undetected error probability $\epsilon$ is specified.
Like that of the ROVA, the complexity of Hof's modified Viterbi algorithm \cite{Hof_Conv_Bounds_ISCTA_2009} is linear in the blocklength and exponential in the constraint length. The algorithms are functionally equivalent, though the ROVA computes $\mathrm{P}(Y^n)$ in order to yield $\mathrm{P}(\hat{X}^n|Y^n)$, while Hof's algorithm computes $\sum_{m' \neq m} \mathrm{P}(Y^n|X_{m'}^n)$.

%
%
\subsection{Throughput and Latency}

Bounds on the maximum rate in \cite{Polyanskiy_IT_2011_NonAsym} assume that an infinite-length codebook is available, i.e., that an infinite number of coded symbols may be transmitted until the receiver makes a reliable decoding decision. While such infinite-length code constructions exist (e.g., rateless fountain codes), we demonstrate that the combination of decoding after every symbol and ACK/NACK feedback controlling additional transmissions delivers high rates at short blocklengths even when the codewords have finite length.  Similar behavior is seen in \cite{Chen_ISIT_2013}, which explores the effect of finite-length codewords on the achievable rates of VLFT coding. 

We encode a message with $k = \log M$ message bits into a mother codeword of length $N$. 
The initial transmission is accomplished by pseudo-random rate-compatible puncturing of the mother code \cite{Hagenauer_TCOM_1988}, such that one symbol is transmitted at a time and the receiver uses all received symbols to decode. 

If the receiver requests additional redundancy after the $N$ symbols have been exhausted, the transmitter begins resending the original sequence of symbols and decoding starts from scratch.  While some benefit can be accrued by retaining the $N$ already transmitted symbols (for example, by Chase code combining), we do not exploit this opportunity in our simulations for simplicity.

The latency $\ell$ (i.e., average number of channel uses, or average blocklength) and the throughput $R_t$ of the proposed scheme are given by 
	\begin{IEEEeqnarray}{rCl}
	\ell &=& \frac{1 + \sum\limits_{i=1}^{N-1} P_{\text{NACK}}(i)}{1 - P_{\text{NACK}}(N)}, 
	\label{eqn:lambda_ROVA} \\
	R_t &=& \frac{k}{\ell} (1 - P_{\text{UE}}),
	\label{eqn:Rt_ROVA}
	\end{IEEEeqnarray}
where $P_{\text{NACK}}(i)$ is the probability of  a NACK generated because the ROVA computed the probability of error to be larger than $\epsilon$ when $i$  coded symbols have been received, and $P_{\text{UE}}$ is the overall probability of undetected error.  Both $P_{\text{NACK}}(i)$  and $P_{\text{UE}}$ depend on the target $\epsilon$.\footnote{We obtain $P_{\text{NACK}}(i)$ and $P_{\text{UE}}$ via simulation.} We have included the factor $(1 - P_{\text{UE}})$ in the throughput expression to emphasize that we are only counting messages that are decoded successfully at the receiver (i.e., the goodput).

%
%
\section{Simulation Results}
\label{sec:Simulation_Results}

\subsection{Convolutional Codes}

\begin{table}
\begin{center}
  \caption{Generator polynomials $g_1$, $g_2$, and $g_3$ corresponding to the rate $1$/$3$ convolutional codes used in the ROVA simulations. $d_{\text{free}}$ is the free distance, $A_{d_{\text{free}}}$ is the number of codewords with weight $d_{\text{free}}$, and $L_D$ is the analytic traceback depth.} 
\begin{tabular}{ c | c | c | c | c | c }
  \# memory  & \# states, & \\
  elements, $\nu$ & $s=2^{\nu}$ & $(g_1, g_2, g_3)$  & $d_{\text{free}}$ & $A_{d_{\text {free}}}$ & $L_D$ \\
  \hline \hline
  6 & 64 & (117, 127, 155) & 15 & 3 & 21 \\
  8 & 256 & (575, 623, 727) & 18 & 1 & 25 \\
  10 & 1024 & (2325, 2731, 3747) & 22 & 7 & 34 \\
  \hline
\end{tabular}
\label{tbl:conv_codes}
\end{center}
\vspace{-8pt}
\end{table}

Table \ref{tbl:conv_codes} lists the rate $1$/$3$ convolutional codes from \cite[Table 12.1]{Lin_2004_ECC} that were used as mother codes for simulations. The codes selected have the optimum free distance $d_{\text{free}}$, which is listed along with the analytic traceback depth $L_D$ \cite{Anderson_Traceback_TransIT_1989}. Zero-tail trellis termination is used to facilitate the ROVA, resulting in rate loss (compared to a tail-biting code with no termination bits) at short blocklengths. 
For a code with $\nu$ memory elements and rate $R = \frac{k + \nu}{N}$, the code's effective information rate is $R_{\text{eff}} = \frac{k}{N}$ and the rate loss factor is $\frac{\nu}{k+\nu}$.

As part of the ROVA, the original Viterbi algorithm is used to compute the maximum-likelihood estimate of the transmitted sequence $X^n$ (representing the mother codeword $X^N$) after each new symbol is received. As soon as the estimate is sufficiently reliable, i.e., $\mathrm{P}(\hat X^n|Y^n) \geq 1 - \epsilon$, transmission terminates and a new codeword is simulated.

%
\subsection{Binary Symmetric Channel (BSC)}

\begin{figure}
  \centering
    	\scalebox{0.47}{\includegraphics{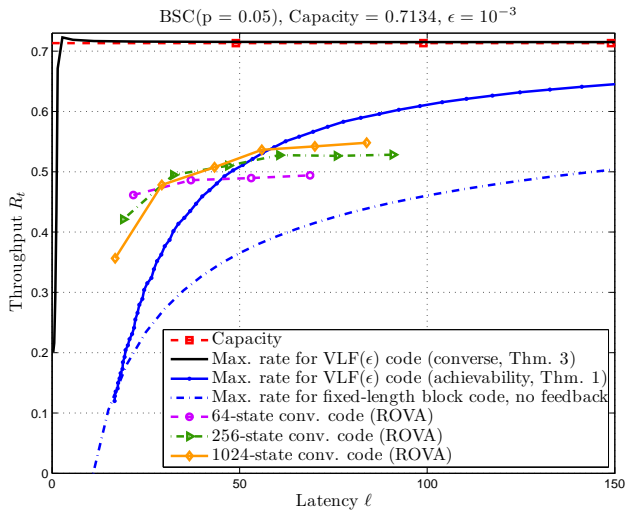}}
    \hspace{0.1in}
    \vspace{-4pt}
\caption{Simulation results of variable-length coding for the BSC using the ROVA as a stopping criterion.}
\label{fig:ROVA_bsc_sims}
  \vspace{-16pt}
\end{figure}

Fig.~\ref{fig:ROVA_bsc_sims} compares the results of VLF simulations using convolutional codes and the ROVA with Polyanskiy et al.'s upper bound and random-coding lower bound on rate for a BSC with crossover probability $p$=$0.05$ and target probability of error $\epsilon$$=$$10^{-3}$. The (asymptotic) capacity of the BSC with crossover probability $p$ is $C_\text{BSC}~=~1~-~h_\text{b}(p)$.
The upper bound is from Thm.~\ref{thm:converse_bsc} and the lower bound is from Thm.~\ref{thm:achievability}. 

Though the upper and lower bounds for VLF codes coincide asymptotically, there is a considerable gap when latency is below 100 bits, a region in which convolutional codes can deliver high rates. At the shortest blocklengths, the 64-state code with the fewest memory elements performs best, due to the trellis-termination rate loss of the codes with larger constraint lengths. However, as the message size $k$ increases (and the latency $\ell$ increases), the larger 1024-state code delivers superior throughput performance. As the latency continues to increase, the codes' throughputs fall below that of the VLF achievability bound.  As the average latency grows, the power of random coding grows but the power of the convolutional codes does not improve significantly once the average latency is beyond twice the traceback depth $L_D$ of the convolutional code \cite{Anderson_Traceback_TransIT_1989}.

Fig. \ref{fig:ROVA_bsc_sims} also includes information-theoretic limits on the maximum rate attainable at short blocklengths without feedback. The ``Max. rate for fixed-length block code, no feedback" curve uses the channel dispersion to compute the maximum rate, which tightly approximates both the achievability and converse bounds when there is no feedback \cite{Polyanskiy_CCR_2010}.

%
\subsection{Additive White Gaussian Noise (AWGN) Channel}

\begin{figure}
  \centering
    	\scalebox{0.47}{\includegraphics{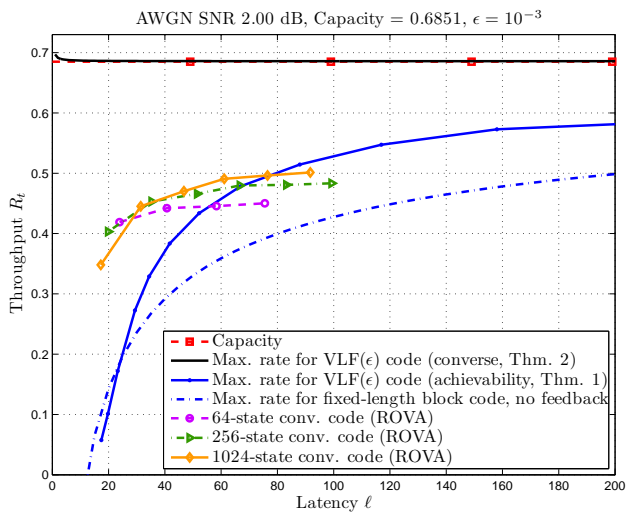}}
    \hspace{0.1in}
\caption{Simulation results of variable-length coding for the AWGN channel using the ROVA as a stopping criterion.}
\label{fig:ROVA_awgn_sims}
  \vspace{-12pt}
\end{figure}

Fig.~\ref{fig:ROVA_awgn_sims} compares the results of VLF simulations using convolutional codes and the ROVA with Polyanskiy et al.'s upper bound and random coding lower bound on rate for the AWGN channel with SNR $2.00$ dB and target $\epsilon$$=$$10^{-3}$.  The (asymptotic) capacity of the AWGN channel with SNR $P$ is $C_\text{AWGN} = \frac{1}{2}\log (1 + P)$. The upper bound is from Thm.~\ref{thm:converse_awgn} and the lower bound is from Thm.~\ref{thm:achievability}, particularized to the Gaussian channel. Details of the VLF($\epsilon$) computation for the AWGN channel are provided in the Appendix.

%
%

The convolutional codes shown in Fig.~\ref{fig:ROVA_awgn_sims} assume transmission over a binary-input AWGN (BI-AWGN) channel (i.e., using BPSK signaling) with soft-decision decoding, which has a maximum Shannon capacity of 1 bit per channel use even when the SNR $P$ is unbounded. However, we have compared the throughput of the ROVA simulations with the capacity $C_\text{AWGN}$ and VLF bounds for the full AWGN channel (i.e., with real-valued inputs drawn i.i.d. $\sim \mathcal{N}(0,P)$).
For SNR=2 dB and capacity 0.6851 this is a minor concern.  However, simulations at higher SNRs would require a higher order modulation (e.g., QAM) to achieve rates above 1. 

%
%

%
%
\section{Discussion and Conclusion}
\label{sec:conc}

For latencies (average blocklengths) on the order of 50 or 75 transmissions and less for the BSC and AWGN channel, respectively, VLF simulations with convolutional coding and ROVA outperform the random coding bound on VLF codes.  Furthermore, every operation performed in the simulation is implementable and only one bit of feedback is required per decoding attempt.   As discussed above, the ROVA for computing the probability of error associated with the most likely codeword has the same order of complexity as the Viterbi algorithm.
  
 \subsection{Grouped Transmissions for Practical Decoding}

For comparison with the VLF random coding bounds, we focused attention to the case where decoding is attempted every symbol.  
Decoding less frequently is  more practical due to the round-trip delay inherent in the feedback loop and because of the complexity of performing the ROVA after each received symbol.  Decoding (with the ROVA) only after groups of symbols (``packet" transmissions) is a natural extension of this work.  
Limiting the frequency of decoding increases latency, but the additional latency can be minimal if the decoding interval is well-chosen, as shown for VLFT in \cite{Chen_ISIT_2013}.   Fricke et al. \cite{Fricke_Reliability_HARQ_TCOM_2009} used this approach in a reliability-based hybrid ARQ scheme, though that work was not focused on the short-blocklength regime.

%
\subsection{On the Use of Convolutional Codes}
\label{sec:note}

Convolutional codes were chosen based on their error-correcting performance at short blocklengths (as compared to, for example,  LDPC and turbo codes) and relatively low decoding complexity.   In our simulations, throughput performance flattens after latency reaches about 75 symbols, consistent with the analytical traceback depths of $L_D\leq34$.  Unlike LDPC codes and turbo codes, the power of a convolutional code does not increase with blocklength; blocklengths greater than two or three times the analytical traceback depth won't significantly lower the error rate and hence won't improve throughput. The achievability of Thm.~\ref{thm:achievability} is still valid; some code with the specified performance exists, but convolutional codes are not likely candidates for achievability at moderate blocklengths.

Recent work by Maiya et al. \cite{Maiya_Costello_Low_Lat_Coding_2012} compared fixed-blocklength convolutional codes and LDPC codes without feedback to determine which codes yielded the best performance at low latencies (not in an incremental redundancy setting).  They showed that for a fixed target error rate (e.g., BER~$=$~$10^{-6}$) and code rate (e.g., $R_c = \frac{1}{2}$), Viterbi-decoded convolutional codes offered the best performance at low latency (e.g., less than 100 bits) and that LDPC codes decoded with iterative message passing offered the best performance for high latencies (e.g., greater than 220 bits). In an intermediate range of latencies (e.g., 100 to 220 bits), convolutional codes with stack sequential decoding were optimal.\footnote{Due to decoding complexity concerns, Viterbi decoding was limited to convolutional codes with 10 or fewer memory elements in \cite{Maiya_Costello_Low_Lat_Coding_2012}.}  This suggests that in order to deliver throughput above the VLF lower bound at moderate blocklengths, LDPC codes or stack sequential decoding of convolutional codes may be suitable.

%
%
\appendix

%
%
\section{VLF($\epsilon$) Codes}
\label{sec:appendix}

\subsection{Information Density of AWGN Channel}
\label{sec:iXY_awgn}

The information density at blocklength $n$, $i(X^n;Y^n)$, for channel input $X^n$ and channel output $Y^n$ is defined  as \cite{Polyanskiy_IT_2011_NonAsym}:
	\begin{IEEEeqnarray}{rCl}
	i(X^n;Y^n) = \log \frac {dP_{Y^n|X^n=x^n} }{dP_{Y^n}}.
	\end{IEEEeqnarray}
For the memoryless AWGN channel with SNR $P$, we have $P_{X^n}$$\sim$${\cal N}(0,P I_n)$, $P_{Z^n}$$\sim$${\cal N}(0,I_n)$, $P_{Y^n}$$\sim$${\cal N}(0,(1+P)I_n)$, and $P_{Y^n|{X^n=x^n}}$$\sim$${\cal N}(x^n,I_n)$, where $I_n$ is the $n \times n$ diagonal matrix. The AWGN information density is:
	\begin{IEEEeqnarray}{rCl}
	i(&X^n&;Y^n) = \log \bigg( (1+P)^{n/2} \frac {e^{-\frac{1}{2} ||y^n - x^n||^2}}{e^{-\frac{1}{2(1+P)} ||y^n||^2}} \bigg) \\
	&=& n C_\text{AWGN} + \frac{\log e}{2} \bigg(-||y^n - x^n||^2 + \frac{||y^n||^2}{(1+P)} \bigg) \\
	&=& n C_\text{AWGN} + \frac{\log e}{2} \sum\limits_{j=1}^{n} \bigg( -(y_j - x_j)^2 + \frac{ y_j^2}{1+P} \bigg) 
	\label{eqn:iXY} \\
	&=& n C_\text{AWGN} + \frac{\log e}{2} \sum\limits_{j=1}^{n} \bigg( -z_j^2 + \frac{(x_j + z_j)^2}{1+P}  \bigg),
	\label{eqn:iXYsum}
	\end{IEEEeqnarray}
where $||y^n||$ is the standard $\ell_2$-norm for Euclidean space.

Due to spherical symmetry, we can write $x^n~$$=$$~(||x^n||,0,0,\dots,0)$ w.l.o.g., which gives us
	\begin{IEEEeqnarray}{rCl}
\vspace{-6pt}
	||x^n + z^n||^2 &=& (x_1 + z_1)^2 + \sum\limits_{j=2}^{n} z_j^2 \nonumber \\
	&=& \big( \sqrt{P \chi_1^2} + Z \big)^2 + \chi_2^2
	\end{IEEEeqnarray}
and $||z^n||^2 = Z^2 + \chi_2^2$, where $Z$$\sim$${\cal N}(0,1)$, $\chi_1^2$ is a chi-square with $n$ degrees of freedom, and $\chi_2^2$ is a chi-square with $(n-1)$ degrees of freedom.
Substituting this into \eqref{eqn:iXYsum}, we have
	\begin{IEEEeqnarray}{rCl}
	i(&X^n&;Y^n) = n C_\text{AWGN} + \frac {\log e}{2} \times \nonumber \\
	& & \bigg(-Z^2 - \frac{P}{1+P} \chi_2^2 + \frac{\left(\sqrt{P \chi_1^2} + Z \right)^2}{1+P}  \bigg).
	\label{eqn:iXY_num}
	\end{IEEEeqnarray}

\subsection{Computation of VLF($\epsilon$) Achievability Curve}
\label{sec:vlf_awgn}

Consider the following bound on $\P[\bar \tau \leq n]$:
	\begin{IEEEeqnarray}{rCl}
	\P[\bar \tau \leq n] 	&=& \E[ \mathrm{1} \{\bar \tau \leq n\}] \\
	&=& \E[ \mathrm{1} \{\tau \leq n\} \exp\{-i(X^{\tau};Y^{\tau}\}] \\
	&\leq& \exp \{-\gamma\},
	\end{IEEEeqnarray}
where the last inequality follows because $i(X^{\tau};Y^{\tau}) \geq \gamma$ by definition in \eqref{eqn:vlf_tau}. Accordingly, we can write the following
	\begin{IEEEeqnarray}{rCl}
	\mathrm{P}[ \bar\tau \leq \tau] &=& \sum \limits_{n=0}^{\infty} \P[\tau = n] \P[\bar \tau \leq n] \\
	&\leq& \sum \limits_{n=0}^{\infty} \P[\tau = n] \exp\{-\gamma\} \\
	&=& \exp\{-\gamma\},
	\end{IEEEeqnarray}
which leads to a looser bound on the error probability:
	\begin{IEEEeqnarray}{rCl}
	\epsilon \leq (M-1) \exp\{-\gamma\}.
	\label{eqn:epsilon_gamma}
	\end{IEEEeqnarray}
For a given error constraint $\epsilon'$, we pick the smallest $\gamma$ such that the right-hand side of \eqref{eqn:epsilon_gamma} is smaller than $\epsilon'$, and then compute the average blocklength $\ell$ as
	\begin{IEEEeqnarray}{rCl}
	\ell \leq \mathrm{E}[\tau] &=& \sum \limits_{n=0}^{\infty} \P[\tau > n] \\
	&\leq& \sum \limits_{n=0}^{\infty} \P[i(X^n;Y^n) < \gamma].
	\label{eqn:i_gamma}
	\end{IEEEeqnarray}

For each value of $n$, the term $\P[i(X^n;Y^n) < \gamma]$ can be evaluated numerically using the expression for $i(X^n;Y^n)$ in \eqref{eqn:iXY_num}. For the AWGN channel, this computation involves a 3-dimensional integral over the random variables $Z$, $\chi_1^2$, and $\chi_2^2$.

%
%
\bibliographystyle{IEEEtran}
{\bibliography{AW_bib}}


\end{document}